\documentclass[aps,pre,twocolumn, groupedaddress]{revtex4}
\usepackage[normalem]{ulem}
\usepackage{float}
\usepackage[usenames]{color}
\usepackage{graphicx}
\usepackage{ulem}
\usepackage{amsfonts}
\usepackage{amsmath}
\usepackage{natbib}

\renewcommand{\d}{{\rm d}}

\begin{document}

\title{Molecular motors stiffen non-affine semiflexible polymer networks}

\author{C. P. Broedersz}
\affiliation{Department of Physics and Astronomy, Vrije Universiteit, Amsterdam, The Netherlands}
\author{F. C. MacKintosh}
\affiliation{Department of Physics and Astronomy, Vrije Universiteit, Amsterdam, The Netherlands}

\date{\today}

\begin{abstract}
Reconstituted filamentous actin networks with myosin motor proteins form active gels, in which motor proteins generate forces that drive the network far from equilibrium. This motor activity can also strongly affect the network elasticity; experiments have shown a dramatic stiffening in \emph{in vitro} networks with molecular motors. Here we study the effects of motor generated forces on the mechanics of simulated 2D networks of athermal stiff filaments.  We show how heterogeneous internal motor stresses can lead to stiffening in networks that are governed by filament bending modes. The motors are modeled as force dipoles that cause muscle like contractions. These contractions ``pull out'' the floppy bending modes in the system, which induces a cross-over to a stiffer stretching dominated regime. Through this mechanism, motors can lead to a nonlinear network response, even when the constituent filaments are themselves purely linear. These results have implications for the mechanics of living cells and suggest new design principles for active biomemetic materials with tunable mechanical properties.
\end{abstract}

\maketitle

The mechanics of living cells is largely governed by the cytoskeleton, a complex assembly of various filamentous proteins. Cross-linked networks of actin filaments form one of the major structural components of the cytoskleton. However, this cytoskeleton is driven far from equilibrium by the action of molecular motors that can generate stresses within the meshwork of filaments\cite{alberts_molecular_2002,brangwynne_cytoplasmic_2008,joanny_active_2009}.  Such motor activity plays a key role in various cellular functions, including morphogenesis, division and locomotion. The nonequilibrium nature of motor activity has been demonstrated in simplified reconstituted filamentous actin networks with myosin motors\cite{mizuno_nonequilibrium_2007,brangwynne_nonequilibrium_2008,bendix_quantitative_2008,koenderink_active_2009,schaller_polar_2010}. Even in the absence of motor proteins, such \emph{in vitro} networks of cytoskeletal filaments already constitute a rich class of soft matter systems that exhibit unusual material properties, including a highly nonlinear elastic response to external stress~\cite{gardel_elastic_2004,storm_nonlinear_2005,wagner_cytoskeletal_2006,bausch_bottom-up_2006,tharmann_viscoelasticity_2007,kasza_nonlinear_2009,broedersz_measurement_2010}. This nonlinear response can be exploited using molecular motors~\cite{mizuno_nonequilibrium_2007,koenderink_active_2009}; the network stiffness can be varied by orders of magnitude, depending on motor activity. A quantitative understanding of such active biological matter poses a challenge for theoretical modeling~\cite{kruse_generic_2005,joanny_hydrodynamic_2007,mackintosh_nonequilibrium_2008,levine_mechanics_2009,liverpool_mechanical_2009, mackintosh_active_2010,joanny_active_2009}. 

The nonlinear mechanical response of reconstituted biopolymer networks in many cases reflects the nonlinear force-extension behavior of the constituting cross-links or filaments~\cite{gardel_elastic_2004,storm_nonlinear_2005,wagner_cytoskeletal_2006,kasza_nonlinear_2009,broedersz_nonlinear_2008}. For such networks, there is both theoretical and experimental evidence that internal stress generation by molecular motors can result in network stiffening in direct analogy to an externally applied uniform stress~\cite{koenderink_active_2009,mizuno_nonequilibrium_2007,mackintosh_nonequilibrium_2008,levine_mechanics_2009,liverpool_mechanical_2009,head_nonlocal_2010}. However, the mechanical response of semiflexble polymers is highly anisotropic and is typically much softer to bending than to stretching. In some cases, this renders the network deformation highly non-affine with most of the energy stored in bending modes~\cite{head_distinct_2003,head_deformation_2003,wilhelm_elasticity_2003,heussinger_floppy_2006, das_effective_2007}. Such non-affinely deforming stiff polymer networks can also exhibit a nonlinear mechanical response, even when the network constituents have a linear force-extension behavior~\cite{onck_alternative_2005,lieleg_mechanics_2007,huisman_three-dimensional_2007,conti_cross-linked_2009}. However, the effects of internal stresses generated by molecular motors in such networks are unknown.

\begin{figure}
\begin{center}
\includegraphics[width=\columnwidth]{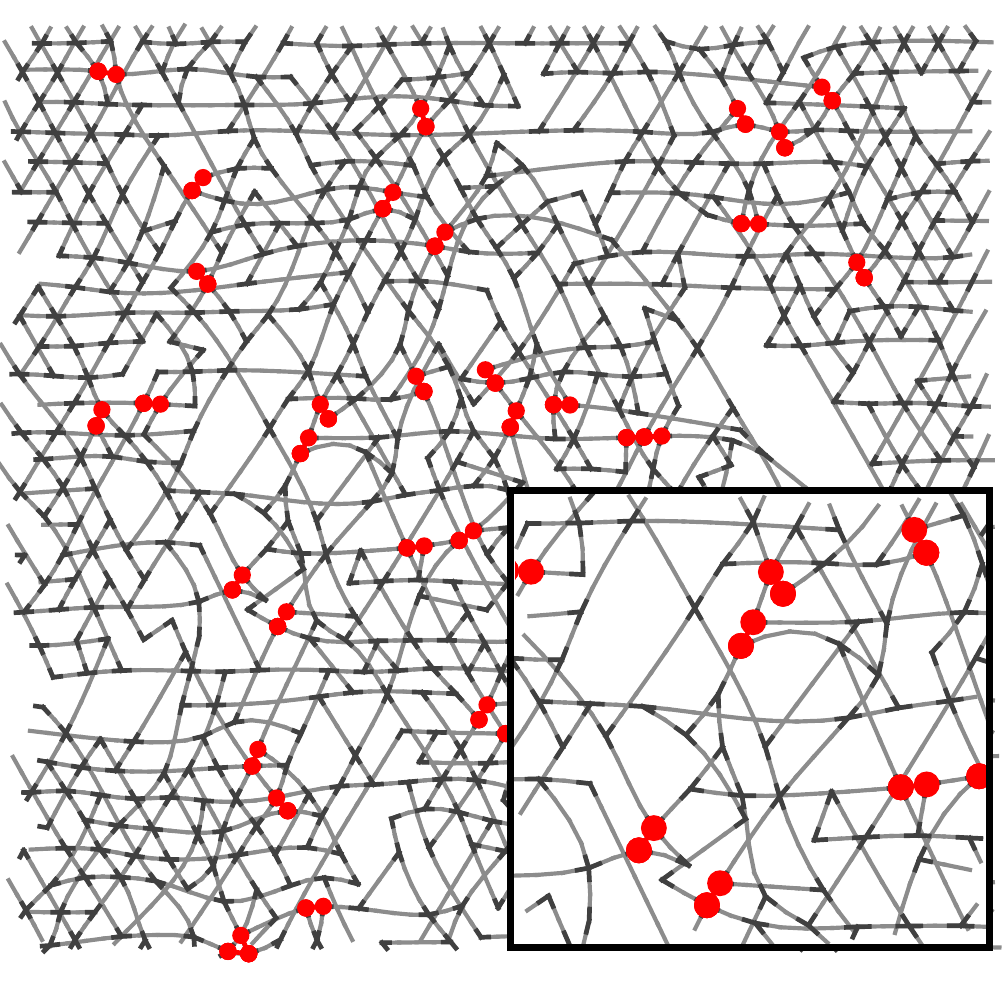}
\caption{\label{fig:network} Example of a portion of the diluted 2D phantom triangular network at $\mathcal{Q}=1/4$ and $\kappa=10^{-3}$. The freely hinging binary cross-links are indicated in black. Motors generate muscle-like contractions, which we model with force dipoles. The segments along which these contractile force dipoles act are indicated with red dumbbells. The inset shows an enlargement of the network.}
\end{center}\vspace{-0.2in}
\end{figure}

Here we study the effects of motor generated forces on the network mechanics in 2D networks of athermal, stiff filaments using simulations. In the absence of motors, these networks can exhibit strain stiffening under an externally applied shear. This behavior has been attributed to a cross-over between two mechanical regimes; at small strains the mechanics is governed by soft bending modes and a non-affine deformation field, while at larger strains the elastic response is governed by the stiffer stretch modes and an affine deformation field\cite{onck_alternative_2005}. We show that motors that generate internal stresses can also stiffen the network. The motors induce force dipoles leading to muscle like contractions, which  "pull out" the floppy bending modes in the system. This induces a cross-over to a stiffer stretching dominated regime. Through this mechanism, motors can lead to network stiffening in non-affine stiff polymer networks in which the constituting filaments in the network are themselves linear elements. These results have implications for the mechanics of living cells and propose new design principles for active biomemetic materials with highly tunable mechanical properties.

\section{The model}

To study the basic effects of internal stress generated by molecular motors on the macroscopic mechanical properties of stiff polymer networks we employ a minimalistic model, which is illustrated in Fig.~\ref{fig:network}.  Filamentous networks in 2D are generated by arranging filaments spanning the system size on a triangular lattice. Since physiological cross-linking proteins typically form binary cross-links, we randomly select two out of the three filaments at every vertex between which we form a binary cross-link. The remaining filament crosses this vertex as a phantom chain, without direct mechanical interactions with the other two filaments. The cross-links themselves hinge freely with no resistance. With this procedure we
can generate disordered \emph{phantom} networks, based on a triangular network, but with local 4-fold ($z=4$) connectivity corresponding to binary cross-links. The use of a triangular lattice avoids, for example, well-known mechanical pathologies of the 4-fold square lattice.
To create quenched disorder in the network, we cut and remove filament segments between vertices with a probability $\mathcal{Q}$. This also has the effect of shortening the filaments. 

The filaments in the network are described by an extensible wormlike chain (EWLC) model with an energy
\begin{equation}\label{eq:H0} 
\mathcal{H}=\frac{1}{2}\kappa \int \d s \left(\frac{\d \hat{t}}{\d s}\right)^2+\frac{1}{2}\mu \int \d s \left(\frac{\d \ell(s)}{\d s}\right)^2,
\end{equation}
where $\kappa$ is the bending rigidity, $\hat{\bf t}$ is the tangent vector at a position $s$ along the polymer backbone and $\frac{\d \ell (s)}{\d s} $ is the local relative change in contour length, or longitudinal strain. We can quantify the relative importance of the stretch and bend contributions by the lengthscale $\ell_b=\sqrt{\kappa/\mu}$; this length scale forms one of the key control parameters for the network mechanics. For simple cylindrical beams with a radius $r$, the stretch modulus $\mu$ is related to $\kappa$ through $\mu_{\rm mech}=4 \kappa/r^2$, and $\ell_b=r/2$. In contrast, a thermally fluctuating semiflexible polymer segment cross-linked in a network on a length-scale $\ell_c$ also has an entropic thermal stretch modulus $\mu_{\rm th}=90 \kappa^2/k_{\rm B} T\ell_c^3$~\cite{mackintosh_elasticity_1995}, where $k_B$ is Boltzmann's constant and $T$ is the temperature. In this case, $\ell_b=\ell_c \sqrt{\ell_c/90 \ell_p}$, where $\ell_p=\kappa/k_{\rm B} T$ is the persistence length. The most relevant values of $\ell_b/\ell_c$ for biopolymer systems range from $10^{-2} - 10^{-1}$.  This range extends from relatively stiff actin filaments to the more flexible intermediate filaments. Various actin binding proteins are capable of forming tightly coupled stiff bundles of actin filaments, which further reduces $\ell_b$. The mechanical and thermal moduli add as springs in series and the total modulus is given by $\mu^{-1}=\mu_{\rm mech}^{-1}+\mu_{\rm th}^{-1}$. 
In the remainder of this paper all lengths are determined in units of the distance between lattice vertices $\ell_0$ and the bending rigidity $\kappa$ is measured in units of $\mu \ell_0^2$. Here, we focus on nonlinearities arising in networks of purely linear elements. Thus, we do not include intrinsic nonlinearities associated with the force-extension curve of thermal filaments. This has been examined theoretically in Refs.~\cite{mackintosh_nonequilibrium_2008,levine_mechanics_2009,liverpool_mechanical_2009}

In our numerical simulations we use a discretized version of Eq.~(\ref{eq:H0}) with a node at and between every lattice vertex. The mid-node allows us to capture buckling down to the single segment length-scale. To model the effect of muscle like contractions induced by molecular motors, we introduce force dipoles in the network \cite{mizuno_nonequilibrium_2007,mackintosh_nonequilibrium_2008,levine_mechanics_2009,head_nonlocal_2010}. These force dipoles are randomly placed at neighboring cross-links. The force dipoles $f_{ij}$ only act along existing bonds and, therefore, do not introduce additional constraints in the network. The total energy of the system includes a sum of the EWLC Hamiltonian over all filament segments and the work extracted by the force dipoles
\begin{equation}\label{eq:Enetwork} 
E=\sum_i \mathcal{H}_{i}-\sum_{<ij>}  f_{ij} r_{ij},
\end{equation}
where $r_{ij}$ is the distance between cross-link $i$ and $j$. The force dipoles are numerically implemented by shortening the effective rest length of the bond along which the motors acts in the stretch term of the energy (Eq.~\ref{eq:H0}). The rest length is reduced by an amount $\delta r_{ij}^{(0)}$;
the resulting force  is given by  $\mu \delta r_{ij}^{(0)}/\ell_0 \le \mu$. The effects of internal motor generated stresses modeled in this way is illustrated in Fig.~\ref{fig:network}.

To investigate the mechanical response of the network, an external strain $\gamma$ is applied by translating one of the horizontal boundaries to which the filaments are attached. The internal degrees of freedom of the network are relaxed by minimizing the energy using a conjugate gradient algorithm~\cite{vetterling_numerical_2002}. To reduce edge effects periodic boundary condition are employed at all boundaries. The linear shear modulus of a network of size $W^2$ is related to the energy $G=\frac{2}{W^2}\frac{E}{\gamma^2}$ for small strains. In the nonlinear regime it is common to determine the differential modulus $K=\frac{1}{W^2}\frac{\d^2 E}{\d \gamma^2}$, which reduces to $G$ for small $\gamma$. Similarly, the stress can be calculated in the nonlinear regime through $\sigma_{\rm ext}=\frac{1}{W^2}\frac{\d E}{\d \gamma}$. These measurements allow us quantify the mechanical response of the system. Here we use system sizes ranging from $W^2\simeq 3000$ to $8000$.

\section{Results and Discussion}
\subsection{Passive networks}

\begin{figure}
\begin{center}
\includegraphics[width=0.9 \columnwidth]{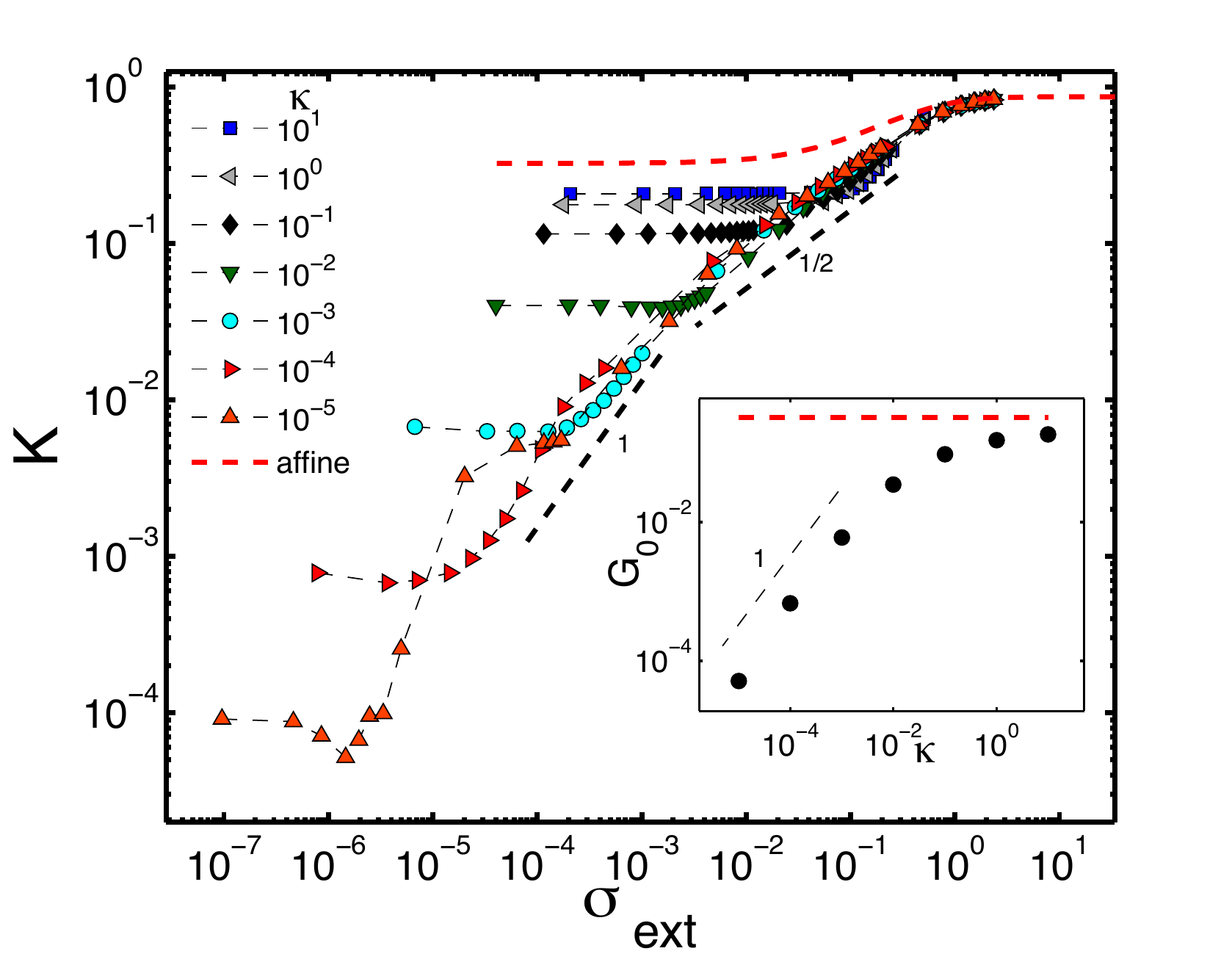}
\caption{\label{fig:NLpassiveK} The differential shear modulus $K=\d \sigma/\d \gamma$ as a function of the applied external stress $\sigma_{\rm ext}$ for various ratios of bending rigidities $\kappa$ and fixed $\mathcal{Q}=1/4$. $K$ and $\sigma_{\rm ext}$ are measured in units of $\mu/\ell_0$. The affine prediction is shown as a red dashed line, which constitutes an upper bound to the elastic response. Although definite powerlaw regimes appear to be absent, the stiffening curves for floppy systems with $\kappa\lesssim 10^{-3}$ initially show a stiffening behavior of approximately $K \sim \sigma$ that crosses over to a regime $K \sim \sigma^{1/2}$ at large stress, as shown by the dashed lines indicating slopes of $1$ and $1/2$.  For stiffer systems with $\kappa\gtrsim 10^{-2}$, only the second of these regimes is apparent. The inset shows the linear shear modulus $G$ as a function of $\kappa$, and the red dashed line indicates the affine prediction.}
\end{center}\vspace{-0.2in}
\end{figure}

\begin{figure}
\begin{center}
\includegraphics[width=0.9 \columnwidth]{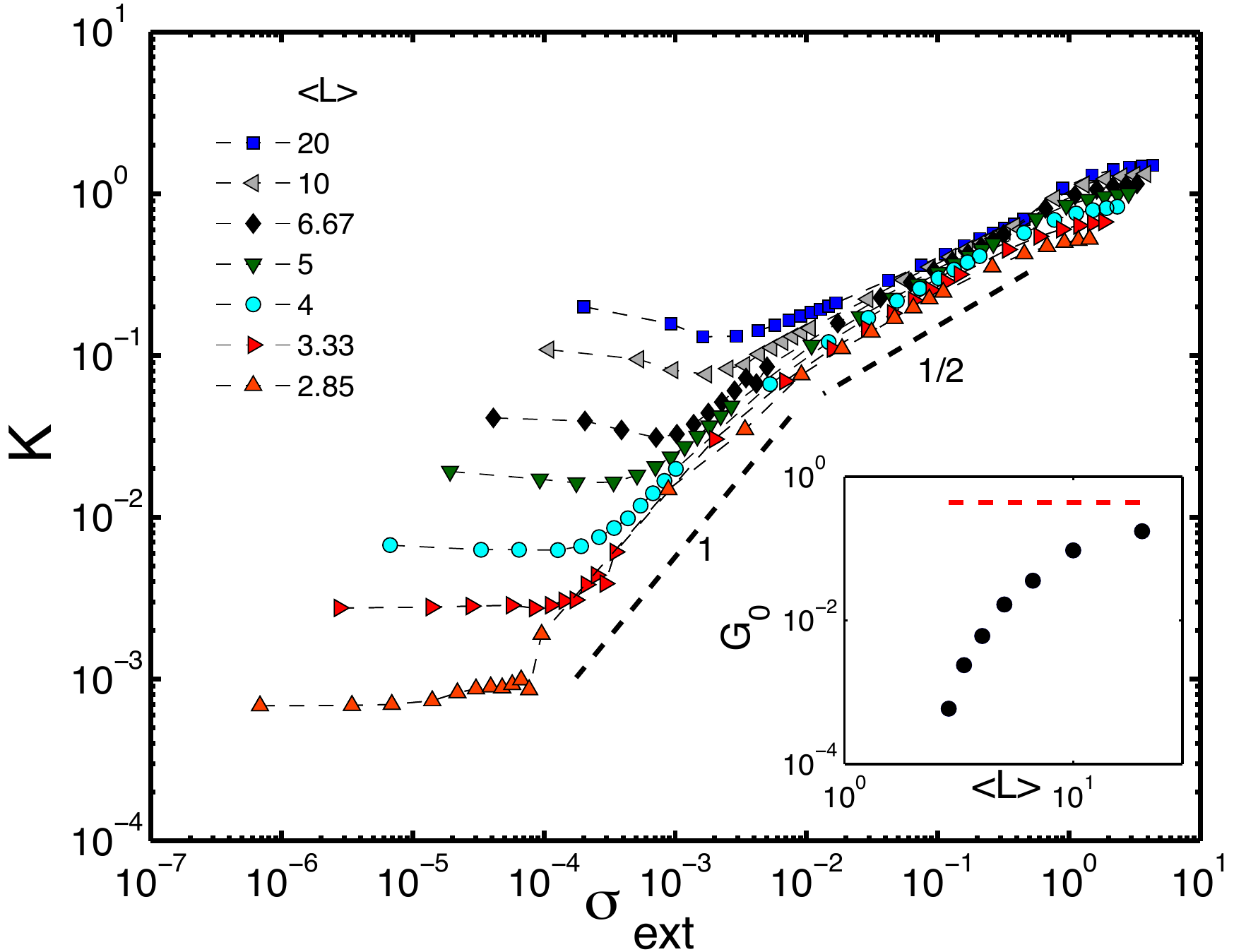}
\caption{\label{fig:NLpassiveP} The differential shear modulus $K=\d \sigma/\d \gamma$ as a function of the applied external stress $\sigma_{\rm ext}$ for various values of $\langle L \rangle$ at fixed bending rigidity $\kappa=10^{-3}$.   $K$ and $\sigma_{\rm ext}$ are measured in units of $\mu/\ell_0$. Although definite powerlaw regimes appear to be absent, the stiffening curves for $\langle L\rangle\lesssim 5$ initially show a stiffening behavior of approximately $K \sim \sigma$ that crosses over to a regime $K \sim \sigma^{1/2}$ at large shear, as shown by the dashed lines that indicate a slope of $1$ and $1/2$. For longer filaments, only the second, weaker stiffening response is apparent. The inset shows the linear shear modulus $G$ as a function of the average filament length $\langle L \rangle$, and the red dashed line indicates the affine prediction in the high molecular weight limit.}
\end{center}\vspace{-0.2in}
\end{figure}
We probe the 2D phantom triangular networks by determining both the linear and nonlinear elastic response of the networks in the absence of motors. The linear mechanical response of diluted networks ($\mathcal{Q}>1$) exhibits two distinct mechanical regimes. At low $\kappa$, the shear modulus $G$ scales directly with $\kappa$, as shown in the inset of Fig.~\ref{fig:NLpassiveK}. This demonstrates that in this regime the macroscopic mechanics is governed by filament bending deformation modes. By contrast, at large $\kappa$ the shear modulus asymptotically approaches a limit in which $G$ is independent of $\kappa$ indicative of a stretching dominated regime. These result are consistent with previous observations on 2D mikado networks~\cite{head_distinct_2003,head_deformation_2003,wilhelm_elasticity_2003}. 

These mechanical regimes have important implications for the nonlinear elastic response. When a large external shear is imposed on a network that is initially in the bending dominated regime, the differential modulus $K=\frac{\d \sigma}{\d \gamma}$ increases strongly as a function of external stress $\sigma_{\rm ext}$, as shown in Fig.~\ref{fig:NLpassiveK}. Previous studies have observed similar stiffening in networks with strictly linear elements~\cite{onck_alternative_2005,lieleg_mechanics_2007,huisman_three-dimensional_2007,conti_cross-linked_2009}. This remarkable behavior has been explained in terms of a \emph{strain}-induced cross-over from a bending to a stretching dominated regime. At low stresses the network mechanics is governed by bending modes, which for small $\kappa$ constitute the softest modes in the system. However, when the stress is increased the deformations become correspondingly large and the stretching of filaments is no longer avoidable. This picture is consistent with our simulations. When a substantial shear is imposed the stiffening curves---over a large range of bending rigidities---converge to a single curve that is consistent with the affine prediction, shown as a red dashed line in Fig.~\ref{fig:NLpassiveK}. This calculation also demonstates that even an affinely deforming network of strictly linear elements stiffens under shear. This stiffening behavior is purely due to geometric effects; under shear the network becomes increasingly anisotropic and the filaments reorient to line up in the shear direction~\cite{broedersz_effective-medium_2009}. The extent of this purely geometric stiffening is, however, limited, as can be seen in the figure. Moreover, this geometrically-stiffened limit represents an upper bound on the stiffness of networks with purely linear elements. Such systems cannot stiffen indefinitely. 

In addition to $\kappa$, the average length of filaments in the system $\langle L\rangle$ constitutes an important control parameter for the linear response. We can probe this by varying $\mathcal{Q}$, since the average length of filaments is given by $\langle L\rangle=1/\mathcal{Q}$~\cite{broedersz_future_2010}. Consistent with previous work~\cite{head_distinct_2003,head_deformation_2003,wilhelm_elasticity_2003}, a cross-over from a non-affine bending regime and an affine stretching regime can also be achieved by increasing $\langle L\rangle$, as shown in the inset of Fig.~\ref{fig:NLpassiveP}. In the high molecular weight limit, $\langle L\rangle\rightarrow\infty$, the system responds purely affinely. We estimate that in experimental biopolymer systems $\langle L\rangle$ varies a over a range of order 5-30, in units of the network mesh size. The strong dependence of the linear elastic response on $\langle L \rangle$ is also reflected in the nonlinear response (Fig.~\ref{fig:NLpassiveP}). Networks with shorter filaments are increasingly governed by soft bending modes and thus exhibit a greater degree of stiffening under shear.

In the absence of motors, we find that our diluted phantom triangular networks exhibit a linear and nonlinear response to external shear that is consistent with previous work on 2D off-lattice networks of stiff filaments~\cite{head_distinct_2003,head_deformation_2003,wilhelm_elasticity_2003}. Our phantom triangular networks thus provide a good model system to study the effects of internal stresses generated by molecular motors in athermal networks. 

\subsection{Active networks}
\begin{figure}
\begin{center}
\includegraphics[width=\columnwidth]{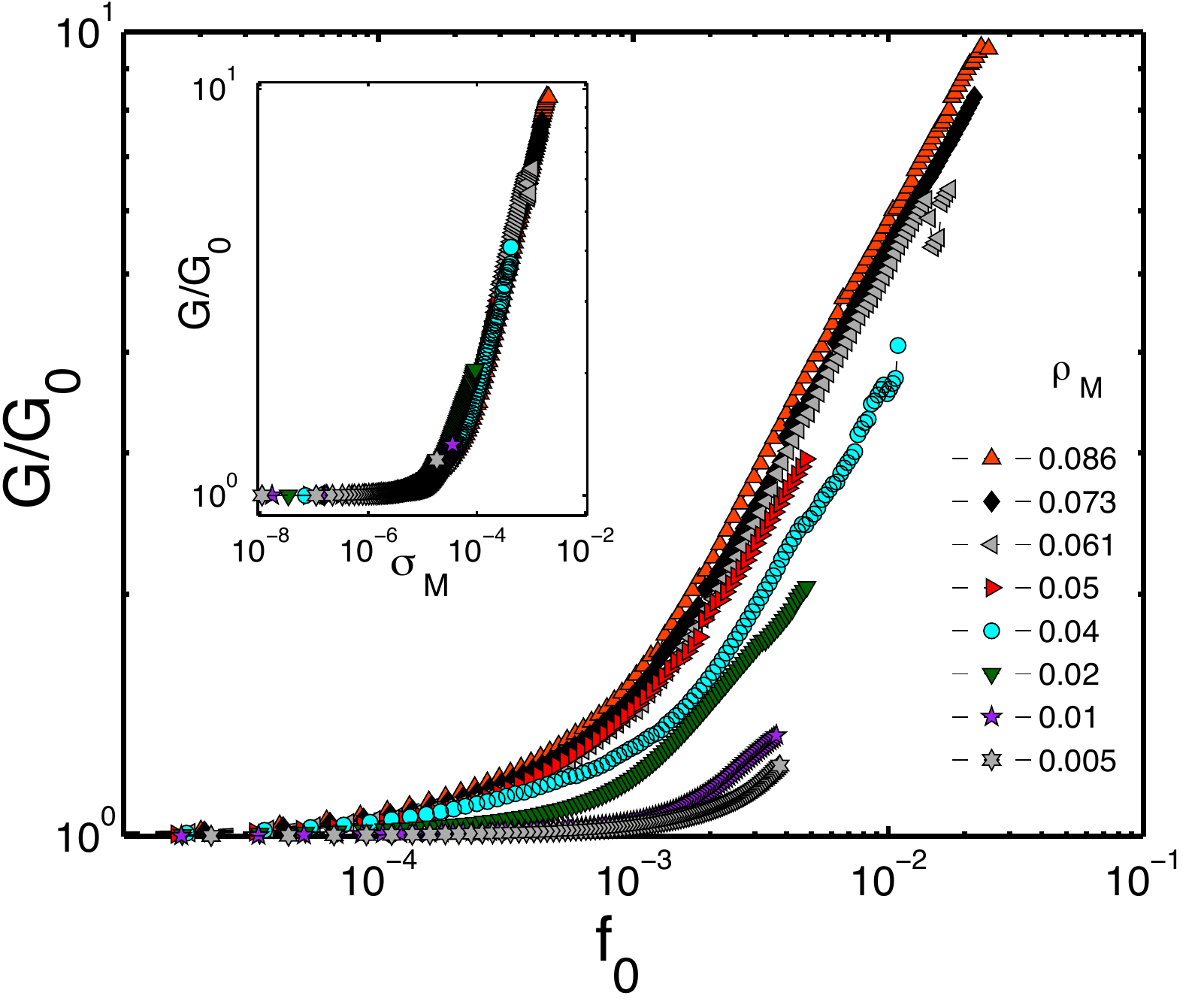}
\caption{\label{fig:Mvaryden} The shear modulus $G$ as a function of force exerted per motor $f_0$ for various motor densities $\rho_{\rm M}$ at fixed $\mathcal{Q}=1/4$ and $\kappa=10^{-3}$. The shear modulus $G$ is normalized by the shear modulus $G_0$ of the passive network. The inset shows the shear modulus $G_0$ as a function of the generated stress $\sigma_{\rm M}$. The apparent collapse of these curves supports supports the hypotheses that $\sigma_{\rm M}$ is the appropriate control variable.}
\end{center}\vspace{-0.2in}
\end{figure}

To investigate the effect of motor generated stresses we introduce force dipoles in the network at various densities $\rho_{\rm M}$. The shear modulus $G$ increases strongly when the force exerted by a single motor $f_0$ is increased beyond a threshold value, as shown in Fig.~\ref{fig:Mvaryden}. Interestingly, the motor forces at which the system becomes nonlinear for low motor densities is close to the buckling force threshold $f_b=\pi^2 \kappa/\ell_c^2\approx2\times10^{-3}$. The buckling force threshold has been identified as an important force-scale for stiffening of these networks under external shear~\cite{conti_cross-linked_2009,onck_alternative_2005}. In addition, these data imply that a minimum motor density is required for motor generated stiffening, consistent with recent experiments~\cite{koenderink_active_2009}. The characteristic motor-generated stress can be expressed as  $\sigma_{\rm M}=\rho_{\rm M} \ell_0 f_0 $. Remarkably, all stiffening curves can be collapsed by expressing the shear modulus as a function of $\sigma_{\rm M}$ (upper inset Fig.~\ref{fig:Mvaryden}). This demonstrates that the characteristic motor generated stress $\sigma_{\rm M}$ is a useful quantity, even though the distribution of stress is likely to be highly heterogenous. 

\begin{figure}
\begin{center}
\includegraphics[width= \columnwidth]{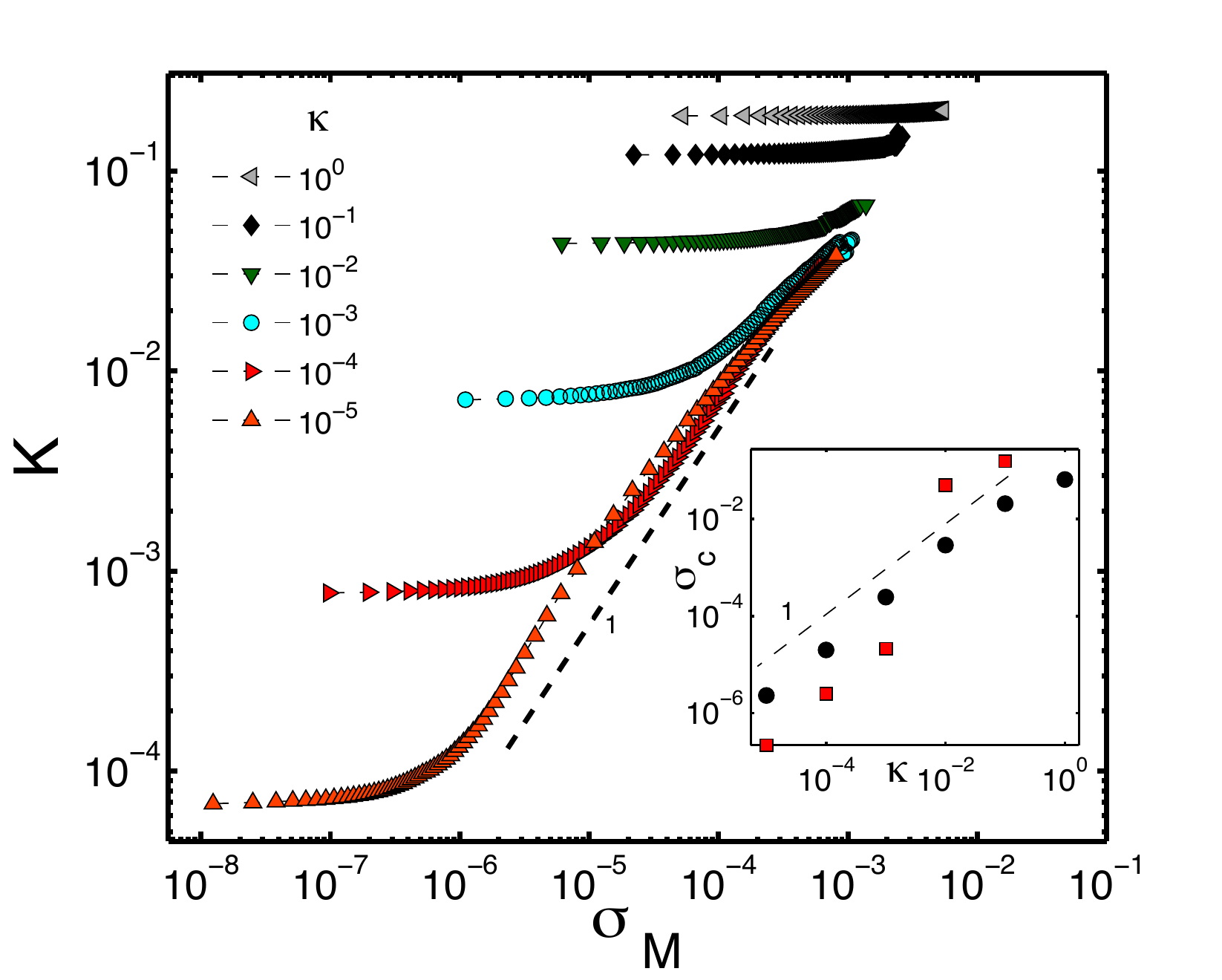}
\caption{\label{fig:Mvarykappa} The linear shear modulus $G$ as a function of motor generated stress $\sigma_{\rm M}$ for various of bending rigidities $\kappa$ at fixed $\mathcal{Q}=1/4$.  The stiffening curves for networks with floppy filaments for $\kappa \lesssim 10^{-3}$ show an approximate scaling behavior given by  $K \sim \sigma$, as shown by the dashed lines that indicate a slope of $1$. The inset shows the critical stress for the onset of stiffening as a function of $\kappa$ for both the active (red squares) and the passive (black circles) systems.}
\end{center}\vspace{-0.2in}
\end{figure}

To explore the nature of the stiffening induced by motors we study the networks' response at various values of $\kappa$. We observe that motor activity dramatically increases the network stiffness  over a range of $\kappa$ values, as shown in Fig.~\ref{fig:Mvarykappa}. Interestingly, the degree of stiffening induced by motors stress is substantially larger for networks with lower $\kappa$, while for large $\kappa$ we observe no stiffening at all. To compare the stiffening between the active and passive networks, we determine the critical stress for the onset of stiffening. When the linear mechanics of the networks is controlled by bending modes ($G\sim\kappa$) we find that $\sigma_c$ scales linearly with $\kappa$ for both active and passive networks, as shown in the inset Fig.~\ref{fig:Mvarykappa}. At larger bending rigidities $\sigma_c$ saturates to a value independent of $\kappa$. Interestingly, the values of $\sigma_c$ for active floppy networks are substantially lower than for the passive networks. This indicates that internally generated motor stress is more effective in network stiffening than an external stress. 

\begin{figure}
\begin{center}
\includegraphics[width=\columnwidth]{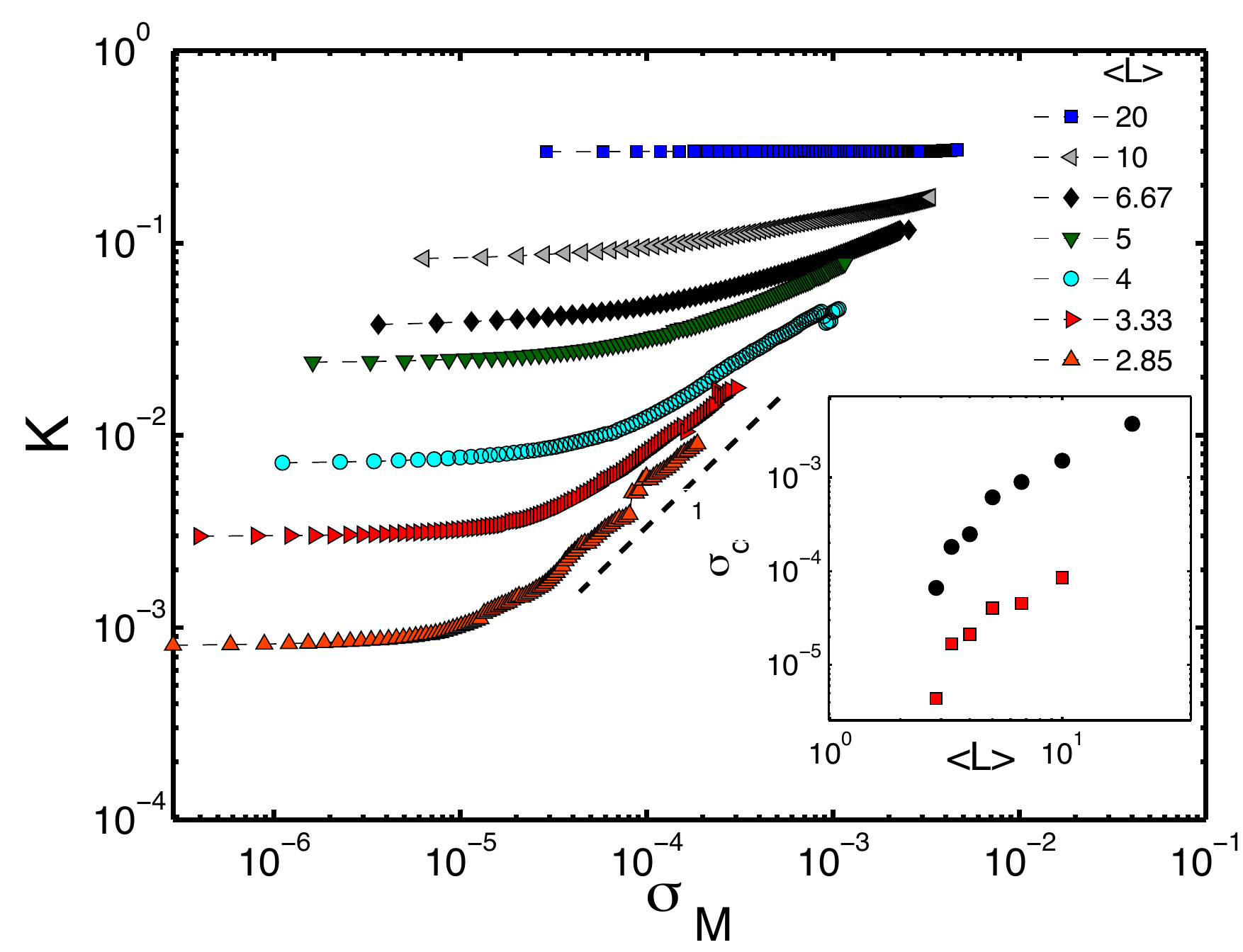}
\caption{\label{fig:MvaryP} The linear shear modulus $G$ as a function of motor generated stress $\sigma_{\rm M}$ for various values of $\langle L\rangle$ at fixes bending rigidity $\kappa=10^{-3}$. The stiffening curves for $\langle L \rangle \lesssim 5$  show an approximate scaling behavior given by $K \sim \sigma$, as shown by the dashed lines that indicate a slope of $1$. The inset shows the critical stress for the onset of stiffening as a function of $\kappa$ for both the active (red squares) and the passive (black circles) systems.}
\end{center}\vspace{-0.2in}
\end{figure}

To identify the role of filament length in motor generated stiffening we vary $\mathcal{Q}$ to tune $\langle L\rangle$. Interestingly, only networks with relatively short filaments stiffen strongly (Fig.~\ref{fig:MvaryP}). Networks with longer filaments are governed increasingly by the stretching modes in the system. This is consistent with the numerical data in Fig.~\ref{fig:Mvarykappa}, for which we observed that only bending dominated networks are capable of stiffening by motor activity.  The critical stress for the onset of stiffening scales in the same way with $\langle L\rangle$ for the active networks as for the passive networks (inset Fig.~\ref{fig:MvaryP}), similar to what we observed for the scaling of $\sigma_c$ with $\kappa$ (inset Fig.~\ref{fig:MvaryP}). Taken together, these results provide evidence that the motor generated stiffening in the active networks derives from the same origin as the stiffening of passive networks under external shear.

\begin{figure}
\begin{center}
\includegraphics[width= \columnwidth]{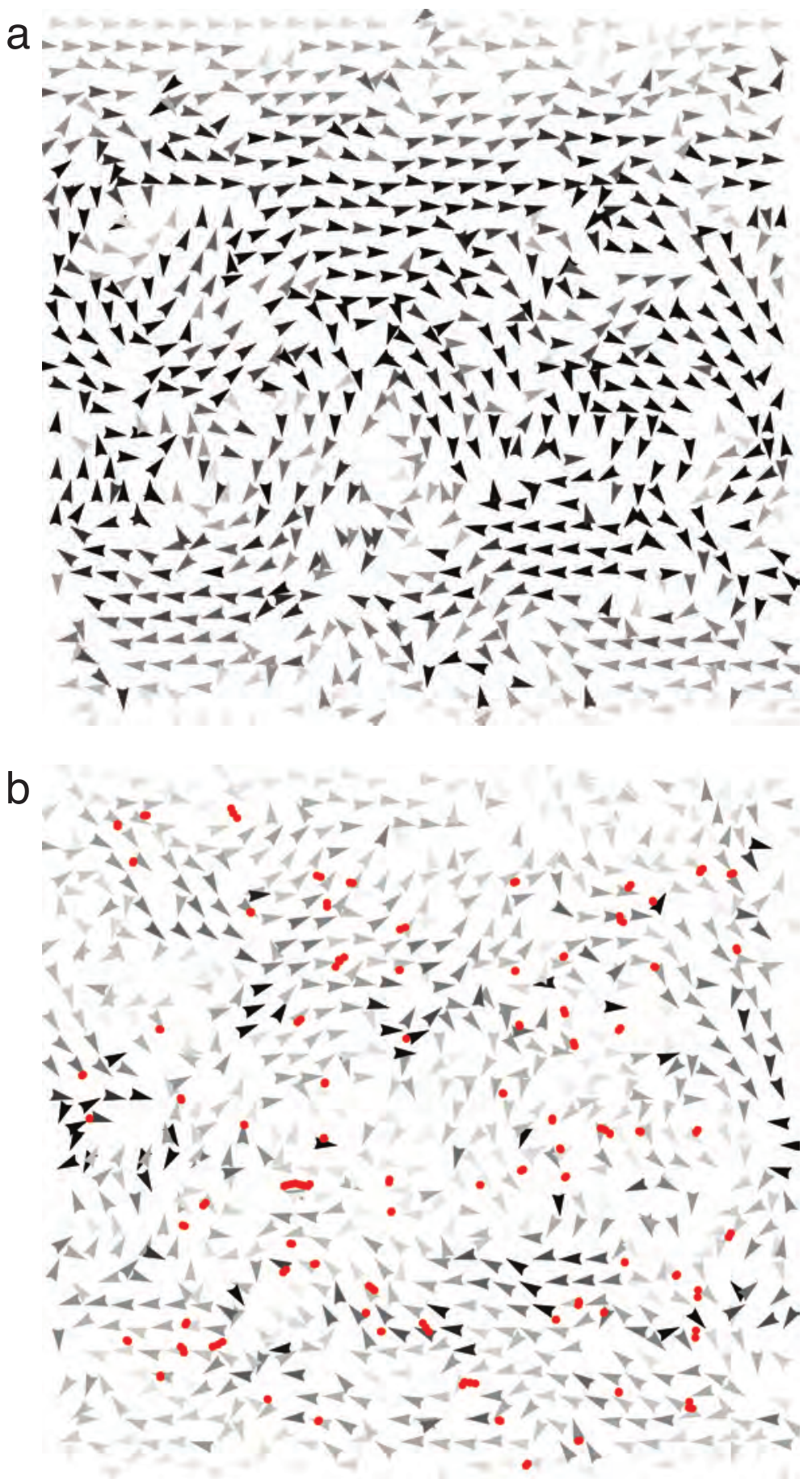}
\caption{\label{fig:NA} The non-affine component of the deformation field under an external shear for a passive (a) network ($\kappa=10^{-3}$ and $\mathcal{Q}=1/4$) and for the same network with motors ($\rho_{\rm M}=0.061$, $f_0\sim10^{-2}$) deep into the stiffened regime. The greyscale of the arrow heads indicate the magnitude of the non-affine deformation; black indicates a large magnitude ($\sim0.01$) and light grey a small magnitude ($\sim0.001$). The motors are shown as red dumbbells.} 
\end{center}\vspace{-0.2in}
\end{figure}

The analogy between external stress and motor generated stress can be further explored by determining the effect of motor activity on the microscopic deformation field. The stiffening in passive networks has been attributed to a shear-induced cross-over between soft bending modes and stiffer stretching modes; concomitant with this cross-over the deformation becomes increasingly affine for larger strains~\cite{onck_alternative_2005}. Our simulations suggest that the same basic mechanism is responsible for the motor generated stiffening in non-affine networks. To further test this picture we investigate the microscopic deformation field of the these networks under a small external shear. We subtract the affine deformation $\delta {\bf r}_i^{({\rm A})}$ of a cross-link $i$ from the actual deformation $\delta {\bf r}_i$ to isolate the non-affine contribution,
\begin{equation}\label{eq:NA} 
\delta {\bf r}_i^{({\rm NA})}=\delta {\bf r}_i-\delta {\bf r}_i^{({\rm A})}
\end{equation}
 Consistent with prior work\cite{head_distinct_2003}  for a passive networks deep in the bending dominated regime, we observe large non-affine deformations, as shown in Fig.~\ref{fig:NA}a. In contrast, when motors are present the non-affine contribution to the deformation field is substantially reduced, as shown in Fig.~\ref{fig:NA}b. Note, that the motors will initially generate highly non-affine deformations and large bends. These results show, however, that the subsequent deformation of this active network under a small external shear is considerably more affine than in the passive case. This provides insight into the motor induced stiffening we observed in our simulations (Figs.~\ref{fig:Mvarykappa} and ~\ref{fig:MvaryP}). Motor activity pulls out the floppy bend modes, which renders the network deformation more affine and, thereby, induces a cross-over from a response governed by bending modes to a response governed by stretching modes.

\section{conclusion}
Here we have show that molecular motors---modeled as force dipoles---stiffen non-affine networks.  Interestingly, we find that only networks that are strongly governed by bending modes are capable of stiffening through motor activity. The internal stresses generated by the motors pull-out the floppy bending modes in the system, leaving the stiff stretching modes. In this way, motors induce a cross-over to a stretching dominated regime, in analogy to prior results on externally-stressed networks~\cite{onck_alternative_2005,conti_cross-linked_2009}. The absence of motor-induced stiffening of our networks in the stretching dominated regime can be attributed to the purely linear force-extension behavior in our model. Analytical studies based on affine \emph{stretching} dominated networks have shown that motor activity can lead to stiffening when the expected non-linear force-extension relation is taken into account~\cite{mackintosh_nonequilibrium_2008,levine_mechanics_2009,liverpool_mechanical_2009}. 

Nevertheless, within the model we consider, with purely linear elements, our results support the qualitative equivalence of external and internal stress in the nonlinear network response~\cite{mizuno_nonequilibrium_2007,koenderink_active_2009,mackintosh_nonequilibrium_2008}. So far, this correspondence has been understood in the context of stretching-dominated networks, with nonlinear filaments~\cite{mackintosh_nonequilibrium_2008,levine_mechanics_2009, liverpool_mechanical_2009}. The present work shows that this analogy is more general. Interestingly, however, there are some quantitative differences between network stiffening by external load vs internal motor stresses. Specifically, our results support the idea that motor stresses can be more effective in generating stiffening, since they act in all directions~\cite{koenderink_active_2009}. By contrast, when a network is externally sheared most stress is focussed on a small fraction of the filaments that are oriented the direction of extension. Furthermore, there are quantitative differences in the form of the stiffening response with stress in the present model. We find that motor contractility leads to an increase in the shear modulus with motor stress $\sigma_{\rm M}$ (Figs.~4, 5) that is approximately given by $G \sim \sigma_{\rm M}^x$, where $x\approx 1$. By contrast, the stiffening by external shear exhibits a more complex dependence on the stress, with two distinct regimes, corresponding to $x\simeq1$ and $x\simeq1/2$. One important difference that sets the passive networks apart, are the geometric effects that arise at large external shears through the collective alignment of filament in the direction of maximum extension.

The results presented here provide further insight into the mechanisms available for the active cellular cytoskeleton to  regulate the mechanical behavior of the cell. Furthermore, these principles can inspire the design of novel active biomemetic materials with tunable elastic properties.
\begin{acknowledgments}
This work was funded in part by FOM/NWO. The authors thank I. Barmes, E. Conti, M. Das and M. Depken for fruitful discussions.
\end{acknowledgments}

\bibliography{broedersz}

\begin{thebibliography}{36}
\expandafter\ifx\csname natexlab\endcsname\relax\def\natexlab#1{#1}\fi
\expandafter\ifx\csname bibnamefont\endcsname\relax
  \def\bibnamefont#1{#1}\fi
\expandafter\ifx\csname bibfnamefont\endcsname\relax
  \def\bibfnamefont#1{#1}\fi
\expandafter\ifx\csname citenamefont\endcsname\relax
  \def\citenamefont#1{#1}\fi
\expandafter\ifx\csname url\endcsname\relax
  \def\url#1{\texttt{#1}}\fi
\expandafter\ifx\csname urlprefix\endcsname\relax\def\urlprefix{URL }\fi
\providecommand{\bibinfo}[2]{#2}
\providecommand{\eprint}[2][]{\url{#2}}

\bibitem[{\citenamefont{Alberts et~al.}(2002)\citenamefont{Alberts, Johnson,
  Lewis, Raff, Roberts, and Walter}}]{alberts_molecular_2002}
\bibinfo{author}{\bibfnamefont{B.}~\bibnamefont{Alberts}},
  \bibinfo{author}{\bibfnamefont{A.}~\bibnamefont{Johnson}},
  \bibinfo{author}{\bibfnamefont{J.}~\bibnamefont{Lewis}},
  \bibinfo{author}{\bibfnamefont{M.}~\bibnamefont{Raff}},
  \bibinfo{author}{\bibfnamefont{K.}~\bibnamefont{Roberts}}, \bibnamefont{and}
  \bibinfo{author}{\bibfnamefont{P.}~\bibnamefont{Walter}},
  \emph{\bibinfo{title}{Molecular Biology of the Cell}}
  (\bibinfo{publisher}{Garland Science}, \bibinfo{year}{2002}),
  \bibinfo{edition}{4th} ed.

\bibitem[{\citenamefont{Brangwynne
  et~al.}(2008{\natexlab{a}})\citenamefont{Brangwynne, Koenderink,
  {MacKintosh}, and Weitz}}]{brangwynne_cytoplasmic_2008}
\bibinfo{author}{\bibfnamefont{C.~P.} \bibnamefont{Brangwynne}},
  \bibinfo{author}{\bibfnamefont{G.~H.} \bibnamefont{Koenderink}},
  \bibinfo{author}{\bibfnamefont{F.~C.} \bibnamefont{{MacKintosh}}},
  \bibnamefont{and} \bibinfo{author}{\bibfnamefont{D.~A.} \bibnamefont{Weitz}},
  \bibinfo{journal}{The Journal of Cell Biology}
  \textbf{\bibinfo{volume}{183}}, \bibinfo{pages}{583}
  (\bibinfo{year}{2008}{\natexlab{a}}).

\bibitem[{\citenamefont{Joanny and Prost}(2009)}]{joanny_active_2009}
\bibinfo{author}{\bibfnamefont{J.}~\bibnamefont{Joanny}} \bibnamefont{and}
  \bibinfo{author}{\bibfnamefont{J.}~\bibnamefont{Prost}},
  \bibinfo{journal}{{HFSP} Journal} \textbf{\bibinfo{volume}{3}},
  \bibinfo{pages}{94} (\bibinfo{year}{2009}).

\bibitem[{\citenamefont{Mizuno et~al.}(2007)\citenamefont{Mizuno, Tardin,
  Schmidt, and {MacKintosh}}}]{mizuno_nonequilibrium_2007}
\bibinfo{author}{\bibfnamefont{D.}~\bibnamefont{Mizuno}},
  \bibinfo{author}{\bibfnamefont{C.}~\bibnamefont{Tardin}},
  \bibinfo{author}{\bibfnamefont{C.~F.} \bibnamefont{Schmidt}},
  \bibnamefont{and} \bibinfo{author}{\bibfnamefont{F.~C.}
  \bibnamefont{{MacKintosh}}}, \bibinfo{journal}{Science}
  \textbf{\bibinfo{volume}{315}}, \bibinfo{pages}{370} (\bibinfo{year}{2007}).

\bibitem[{\citenamefont{Brangwynne
  et~al.}(2008{\natexlab{b}})\citenamefont{Brangwynne, Koenderink,
  {MacKintosh}, and Weitz}}]{brangwynne_nonequilibrium_2008}
\bibinfo{author}{\bibfnamefont{C.~P.} \bibnamefont{Brangwynne}},
  \bibinfo{author}{\bibfnamefont{G.~H.} \bibnamefont{Koenderink}},
  \bibinfo{author}{\bibfnamefont{F.~C.} \bibnamefont{{MacKintosh}}},
  \bibnamefont{and} \bibinfo{author}{\bibfnamefont{D.~A.} \bibnamefont{Weitz}},
  \bibinfo{journal}{Physical Review Letters} \textbf{\bibinfo{volume}{100}},
  \bibinfo{pages}{118104} (\bibinfo{year}{2008}{\natexlab{b}}).

\bibitem[{\citenamefont{Bendix et~al.}(2008)\citenamefont{Bendix, Koenderink,
  Cuvelier, Dogic, Koeleman, Brieher, Field, Mahadevan, and
  Weitz}}]{bendix_quantitative_2008}
\bibinfo{author}{\bibfnamefont{P.}~\bibnamefont{Bendix}},
  \bibinfo{author}{\bibfnamefont{G.}~\bibnamefont{Koenderink}},
  \bibinfo{author}{\bibfnamefont{D.}~\bibnamefont{Cuvelier}},
  \bibinfo{author}{\bibfnamefont{Z.}~\bibnamefont{Dogic}},
  \bibinfo{author}{\bibfnamefont{B.}~\bibnamefont{Koeleman}},
  \bibinfo{author}{\bibfnamefont{W.}~\bibnamefont{Brieher}},
  \bibinfo{author}{\bibfnamefont{C.}~\bibnamefont{Field}},
  \bibinfo{author}{\bibfnamefont{L.}~\bibnamefont{Mahadevan}},
  \bibnamefont{and} \bibinfo{author}{\bibfnamefont{D.}~\bibnamefont{Weitz}},
  \bibinfo{journal}{Biophysical Journal} \textbf{\bibinfo{volume}{94}},
  \bibinfo{pages}{3126} (\bibinfo{year}{2008}).

\bibitem[{\citenamefont{Koenderink et~al.}(2009)\citenamefont{Koenderink,
  Dogic, Nakamura, Bendix, {MacKintosh}, Hartwig, Stossel, and
  Weitz}}]{koenderink_active_2009}
\bibinfo{author}{\bibfnamefont{G.~H.} \bibnamefont{Koenderink}},
  \bibinfo{author}{\bibfnamefont{Z.}~\bibnamefont{Dogic}},
  \bibinfo{author}{\bibfnamefont{F.}~\bibnamefont{Nakamura}},
  \bibinfo{author}{\bibfnamefont{P.~M.} \bibnamefont{Bendix}},
  \bibinfo{author}{\bibfnamefont{F.~C.} \bibnamefont{{MacKintosh}}},
  \bibinfo{author}{\bibfnamefont{J.~H.} \bibnamefont{Hartwig}},
  \bibinfo{author}{\bibfnamefont{T.~P.} \bibnamefont{Stossel}},
  \bibnamefont{and} \bibinfo{author}{\bibfnamefont{D.~A.} \bibnamefont{Weitz}},
  \bibinfo{journal}{Proceedings of the National Academy of Sciences}
  \textbf{\bibinfo{volume}{106}}, \bibinfo{pages}{15192 }
  (\bibinfo{year}{2009}).

\bibitem[{\citenamefont{Schaller et~al.}(2010)\citenamefont{Schaller, Weber,
  Semmrich, Frey, and Bausch}}]{schaller_polar_2010}
\bibinfo{author}{\bibfnamefont{V.}~\bibnamefont{Schaller}},
  \bibinfo{author}{\bibfnamefont{C.}~\bibnamefont{Weber}},
  \bibinfo{author}{\bibfnamefont{C.}~\bibnamefont{Semmrich}},
  \bibinfo{author}{\bibfnamefont{E.}~\bibnamefont{Frey}}, \bibnamefont{and}
  \bibinfo{author}{\bibfnamefont{A.~R.} \bibnamefont{Bausch}},
  \bibinfo{journal}{Nature} \textbf{\bibinfo{volume}{467}}, \bibinfo{pages}{73}
  (\bibinfo{year}{2010}).

\bibitem[{\citenamefont{Gardel et~al.}(2004)\citenamefont{Gardel, Shin,
  {MacKintosh}, Mahadevan, Matsudaira, and Weitz}}]{gardel_elastic_2004}
\bibinfo{author}{\bibfnamefont{M.~L.} \bibnamefont{Gardel}},
  \bibinfo{author}{\bibfnamefont{J.~H.} \bibnamefont{Shin}},
  \bibinfo{author}{\bibfnamefont{F.~C.} \bibnamefont{{MacKintosh}}},
  \bibinfo{author}{\bibfnamefont{L.}~\bibnamefont{Mahadevan}},
  \bibinfo{author}{\bibfnamefont{P.}~\bibnamefont{Matsudaira}},
  \bibnamefont{and} \bibinfo{author}{\bibfnamefont{D.~A.} \bibnamefont{Weitz}},
  \bibinfo{journal}{Science} \textbf{\bibinfo{volume}{304}},
  \bibinfo{pages}{1301} (\bibinfo{year}{2004}).

\bibitem[{\citenamefont{Storm et~al.}(2005)\citenamefont{Storm, Pastore,
  {MacKintosh}, Lubensky, and Janmey}}]{storm_nonlinear_2005}
\bibinfo{author}{\bibfnamefont{C.}~\bibnamefont{Storm}},
  \bibinfo{author}{\bibfnamefont{J.~J.} \bibnamefont{Pastore}},
  \bibinfo{author}{\bibfnamefont{F.~C.} \bibnamefont{{MacKintosh}}},
  \bibinfo{author}{\bibfnamefont{T.~C.} \bibnamefont{Lubensky}},
  \bibnamefont{and} \bibinfo{author}{\bibfnamefont{P.~A.}
  \bibnamefont{Janmey}}, \bibinfo{journal}{Nature}
  \textbf{\bibinfo{volume}{435}}, \bibinfo{pages}{191} (\bibinfo{year}{2005}).

\bibitem[{\citenamefont{Wagner et~al.}(2006)\citenamefont{Wagner, Tharmann,
  Haase, Fischer, and Bausch}}]{wagner_cytoskeletal_2006}
\bibinfo{author}{\bibfnamefont{B.}~\bibnamefont{Wagner}},
  \bibinfo{author}{\bibfnamefont{R.}~\bibnamefont{Tharmann}},
  \bibinfo{author}{\bibfnamefont{I.}~\bibnamefont{Haase}},
  \bibinfo{author}{\bibfnamefont{M.}~\bibnamefont{Fischer}}, \bibnamefont{and}
  \bibinfo{author}{\bibfnamefont{A.~R.} \bibnamefont{Bausch}},
  \bibinfo{journal}{Proceedings of the National Academy of Sciences}
  \textbf{\bibinfo{volume}{103}}, \bibinfo{pages}{13974}
  (\bibinfo{year}{2006}).

\bibitem[{\citenamefont{Bausch and Kroy}(2006)}]{bausch_bottom-up_2006}
\bibinfo{author}{\bibfnamefont{A.~R.} \bibnamefont{Bausch}} \bibnamefont{and}
  \bibinfo{author}{\bibfnamefont{K.}~\bibnamefont{Kroy}}, \bibinfo{journal}{Nat
  Phys} \textbf{\bibinfo{volume}{2}}, \bibinfo{pages}{231}
  (\bibinfo{year}{2006}).

\bibitem[{\citenamefont{Tharmann et~al.}(2007)\citenamefont{Tharmann,
  Claessens, and Bausch}}]{tharmann_viscoelasticity_2007}
\bibinfo{author}{\bibfnamefont{R.}~\bibnamefont{Tharmann}},
  \bibinfo{author}{\bibfnamefont{M.~M. A.~E.} \bibnamefont{Claessens}},
  \bibnamefont{and} \bibinfo{author}{\bibfnamefont{A.~R.}
  \bibnamefont{Bausch}}, \bibinfo{journal}{Physical Review Letters}
  \textbf{\bibinfo{volume}{98}}, \bibinfo{pages}{088103}
  (\bibinfo{year}{2007}).

\bibitem[{\citenamefont{Kasza et~al.}(2009)\citenamefont{Kasza, Koenderink,
  Lin, Broedersz, Messner, Nakamura, Stossel, {MacKintosh}, and
  Weitz}}]{kasza_nonlinear_2009}
\bibinfo{author}{\bibfnamefont{K.~E.} \bibnamefont{Kasza}},
  \bibinfo{author}{\bibfnamefont{G.~H.} \bibnamefont{Koenderink}},
  \bibinfo{author}{\bibfnamefont{Y.~C.} \bibnamefont{Lin}},
  \bibinfo{author}{\bibfnamefont{C.~P.} \bibnamefont{Broedersz}},
  \bibinfo{author}{\bibfnamefont{W.}~\bibnamefont{Messner}},
  \bibinfo{author}{\bibfnamefont{F.}~\bibnamefont{Nakamura}},
  \bibinfo{author}{\bibfnamefont{T.~P.} \bibnamefont{Stossel}},
  \bibinfo{author}{\bibfnamefont{F.~C.} \bibnamefont{{MacKintosh}}},
  \bibnamefont{and} \bibinfo{author}{\bibfnamefont{D.~A.} \bibnamefont{Weitz}},
  \bibinfo{journal}{Physical Review E} \textbf{\bibinfo{volume}{79}},
  \bibinfo{pages}{041928} (\bibinfo{year}{2009}).

\bibitem[{\citenamefont{Broedersz et~al.}(2010)\citenamefont{Broedersz, Kasza,
  Jawerth, Münster, Weitz, and {MacKintosh}}}]{broedersz_measurement_2010}
\bibinfo{author}{\bibfnamefont{C.~P.} \bibnamefont{Broedersz}},
  \bibinfo{author}{\bibfnamefont{K.~E.} \bibnamefont{Kasza}},
  \bibinfo{author}{\bibfnamefont{L.~M.} \bibnamefont{Jawerth}},
  \bibinfo{author}{\bibfnamefont{S.}~\bibnamefont{Münster}},
  \bibinfo{author}{\bibfnamefont{D.~A.} \bibnamefont{Weitz}}, \bibnamefont{and}
  \bibinfo{author}{\bibfnamefont{F.~C.} \bibnamefont{{MacKintosh}}},
  \bibinfo{journal}{Soft Matter}  (\bibinfo{year}{2010}).

\bibitem[{\citenamefont{Kruse et~al.}(2005)\citenamefont{Kruse, Joanny,
  J�licher, Prost, and Sekimoto}}]{kruse_generic_2005}
\bibinfo{author}{\bibfnamefont{K.}~\bibnamefont{Kruse}},
  \bibinfo{author}{\bibfnamefont{J.~F.} \bibnamefont{Joanny}},
  \bibinfo{author}{\bibfnamefont{F.}~\bibnamefont{J�licher}},
  \bibinfo{author}{\bibfnamefont{J.}~\bibnamefont{Prost}}, \bibnamefont{and}
  \bibinfo{author}{\bibfnamefont{K.}~\bibnamefont{Sekimoto}},
  \bibinfo{journal}{The European Physical Journal E}
  \textbf{\bibinfo{volume}{16}}, \bibinfo{pages}{5} (\bibinfo{year}{2005}).

\bibitem[{\citenamefont{Joanny et~al.}(2007)\citenamefont{Joanny, Jülicher,
  Kruse, and Prost}}]{joanny_hydrodynamic_2007}
\bibinfo{author}{\bibfnamefont{J.~F.} \bibnamefont{Joanny}},
  \bibinfo{author}{\bibfnamefont{F.}~\bibnamefont{Jülicher}},
  \bibinfo{author}{\bibfnamefont{K.}~\bibnamefont{Kruse}}, \bibnamefont{and}
  \bibinfo{author}{\bibfnamefont{J.}~\bibnamefont{Prost}},
  \bibinfo{journal}{New Journal of Physics} \textbf{\bibinfo{volume}{9}},
  \bibinfo{pages}{422} (\bibinfo{year}{2007}).

\bibitem[{\citenamefont{{MacKintosh} and
  Levine}(2008)}]{mackintosh_nonequilibrium_2008}
\bibinfo{author}{\bibfnamefont{F.~C.} \bibnamefont{{MacKintosh}}}
  \bibnamefont{and} \bibinfo{author}{\bibfnamefont{A.~J.}
  \bibnamefont{Levine}}, \bibinfo{journal}{Physical Review Letters}
  \textbf{\bibinfo{volume}{100}}, \bibinfo{pages}{018104}
  (\bibinfo{year}{2008}).

\bibitem[{\citenamefont{Levine and {MacKintosh}}(2009)}]{levine_mechanics_2009}
\bibinfo{author}{\bibfnamefont{A.~J.} \bibnamefont{Levine}} \bibnamefont{and}
  \bibinfo{author}{\bibfnamefont{F.~C.} \bibnamefont{{MacKintosh}}},
  \bibinfo{journal}{The Journal of Physical Chemistry B}
  \textbf{\bibinfo{volume}{113}}, \bibinfo{pages}{3820} (\bibinfo{year}{2009}).

\bibitem[{\citenamefont{Liverpool et~al.}(2009)\citenamefont{Liverpool,
  Marchetti, Joanny, and Prost}}]{liverpool_mechanical_2009}
\bibinfo{author}{\bibfnamefont{T.~B.} \bibnamefont{Liverpool}},
  \bibinfo{author}{\bibfnamefont{M.~C.} \bibnamefont{Marchetti}},
  \bibinfo{author}{\bibfnamefont{J.}~\bibnamefont{Joanny}}, \bibnamefont{and}
  \bibinfo{author}{\bibfnamefont{J.}~\bibnamefont{Prost}},
  \bibinfo{journal}{{EPL} {(Europhysics} Letters)}
  \textbf{\bibinfo{volume}{85}}, \bibinfo{pages}{18007} (\bibinfo{year}{2009}).

\bibitem[{\citenamefont{{MacKintosh} and
  Schmidt}(2010)}]{mackintosh_active_2010}
\bibinfo{author}{\bibfnamefont{F.~C.} \bibnamefont{{MacKintosh}}}
  \bibnamefont{and} \bibinfo{author}{\bibfnamefont{C.~F.}
  \bibnamefont{Schmidt}}, \bibinfo{journal}{Current Opinion in Cell Biology}
  \textbf{\bibinfo{volume}{22}}, \bibinfo{pages}{29} (\bibinfo{year}{2010}).

\bibitem[{\citenamefont{Broedersz et~al.}(2008)\citenamefont{Broedersz, Storm,
  and {MacKintosh}}}]{broedersz_nonlinear_2008}
\bibinfo{author}{\bibfnamefont{C.~P.} \bibnamefont{Broedersz}},
  \bibinfo{author}{\bibfnamefont{C.}~\bibnamefont{Storm}}, \bibnamefont{and}
  \bibinfo{author}{\bibfnamefont{F.~C.} \bibnamefont{{MacKintosh}}},
  \bibinfo{journal}{Physical Review Letters} \textbf{\bibinfo{volume}{101}},
  \bibinfo{pages}{118103} (\bibinfo{year}{2008}).

\bibitem[{\citenamefont{Head and Mizuno}(2010)}]{head_nonlocal_2010}
\bibinfo{author}{\bibfnamefont{D.~A.} \bibnamefont{Head}} \bibnamefont{and}
  \bibinfo{author}{\bibfnamefont{D.}~\bibnamefont{Mizuno}},
  \bibinfo{journal}{Physical Review E} \textbf{\bibinfo{volume}{81}},
  \bibinfo{pages}{041910} (\bibinfo{year}{2010}).

\bibitem[{\citenamefont{Head et~al.}(2003{\natexlab{a}})\citenamefont{Head,
  Levine, and {MacKintosh}}}]{head_distinct_2003}
\bibinfo{author}{\bibfnamefont{D.~A.} \bibnamefont{Head}},
  \bibinfo{author}{\bibfnamefont{A.~J.} \bibnamefont{Levine}},
  \bibnamefont{and} \bibinfo{author}{\bibfnamefont{F.~C.}
  \bibnamefont{{MacKintosh}}}, \bibinfo{journal}{Physical Review E}
  \textbf{\bibinfo{volume}{68}}, \bibinfo{pages}{061907}
  (\bibinfo{year}{2003}{\natexlab{a}}).

\bibitem[{\citenamefont{Head et~al.}(2003{\natexlab{b}})\citenamefont{Head,
  Levine, and {MacKintosh}}}]{head_deformation_2003}
\bibinfo{author}{\bibfnamefont{D.~A.} \bibnamefont{Head}},
  \bibinfo{author}{\bibfnamefont{A.~J.} \bibnamefont{Levine}},
  \bibnamefont{and} \bibinfo{author}{\bibfnamefont{F.~C.}
  \bibnamefont{{MacKintosh}}}, \bibinfo{journal}{Physical Review Letters}
  \textbf{\bibinfo{volume}{91}}, \bibinfo{pages}{108102}
  (\bibinfo{year}{2003}{\natexlab{b}}).

\bibitem[{\citenamefont{Wilhelm and Frey}(2003)}]{wilhelm_elasticity_2003}
\bibinfo{author}{\bibfnamefont{J.}~\bibnamefont{Wilhelm}} \bibnamefont{and}
  \bibinfo{author}{\bibfnamefont{E.}~\bibnamefont{Frey}},
  \bibinfo{journal}{Physical Review Letters} \textbf{\bibinfo{volume}{91}},
  \bibinfo{pages}{108103} (\bibinfo{year}{2003}).

\bibitem[{\citenamefont{Heussinger and Frey}(2006)}]{heussinger_floppy_2006}
\bibinfo{author}{\bibfnamefont{C.}~\bibnamefont{Heussinger}} \bibnamefont{and}
  \bibinfo{author}{\bibfnamefont{E.}~\bibnamefont{Frey}},
  \bibinfo{journal}{Physical Review Letters} \textbf{\bibinfo{volume}{97}},
  \bibinfo{pages}{105501} (\bibinfo{year}{2006}).

\bibitem[{\citenamefont{Das et~al.}(2007)\citenamefont{Das, {MacKintosh}, and
  Levine}}]{das_effective_2007}
\bibinfo{author}{\bibfnamefont{M.}~\bibnamefont{Das}},
  \bibinfo{author}{\bibfnamefont{F.~C.} \bibnamefont{{MacKintosh}}},
  \bibnamefont{and} \bibinfo{author}{\bibfnamefont{A.~J.}
  \bibnamefont{Levine}}, \bibinfo{journal}{Physical Review Letters}
  \textbf{\bibinfo{volume}{99}}, \bibinfo{pages}{038101}
  (\bibinfo{year}{2007}).

\bibitem[{\citenamefont{Onck et~al.}(2005)\citenamefont{Onck, Koeman, van
  Dillen, and van~der Giessen}}]{onck_alternative_2005}
\bibinfo{author}{\bibfnamefont{P.~R.} \bibnamefont{Onck}},
  \bibinfo{author}{\bibfnamefont{T.}~\bibnamefont{Koeman}},
  \bibinfo{author}{\bibfnamefont{T.}~\bibnamefont{van Dillen}},
  \bibnamefont{and} \bibinfo{author}{\bibfnamefont{E.}~\bibnamefont{van~der
  Giessen}}, \bibinfo{journal}{Physical Review Letters}
  \textbf{\bibinfo{volume}{95}}, \bibinfo{pages}{178102}
  (\bibinfo{year}{2005}).

\bibitem[{\citenamefont{Lieleg et~al.}(2007)\citenamefont{Lieleg, Claessens,
  Heussinger, Frey, and Bausch}}]{lieleg_mechanics_2007}
\bibinfo{author}{\bibfnamefont{O.}~\bibnamefont{Lieleg}},
  \bibinfo{author}{\bibfnamefont{M.~M. A.~E.} \bibnamefont{Claessens}},
  \bibinfo{author}{\bibfnamefont{C.}~\bibnamefont{Heussinger}},
  \bibinfo{author}{\bibfnamefont{E.}~\bibnamefont{Frey}}, \bibnamefont{and}
  \bibinfo{author}{\bibfnamefont{A.~R.} \bibnamefont{Bausch}},
  \bibinfo{journal}{Physical Review Letters} \textbf{\bibinfo{volume}{99}},
  \bibinfo{pages}{088102} (\bibinfo{year}{2007}).

\bibitem[{\citenamefont{Huisman et~al.}(2007)\citenamefont{Huisman, van Dillen,
  Onck, and der Giessen}}]{huisman_three-dimensional_2007}
\bibinfo{author}{\bibfnamefont{E.~M.} \bibnamefont{Huisman}},
  \bibinfo{author}{\bibfnamefont{T.}~\bibnamefont{van Dillen}},
  \bibinfo{author}{\bibfnamefont{P.~R.} \bibnamefont{Onck}}, \bibnamefont{and}
  \bibinfo{author}{\bibfnamefont{E.~V.} \bibnamefont{der Giessen}},
  \bibinfo{journal}{Physical Review Letters} \textbf{\bibinfo{volume}{99}},
  \bibinfo{pages}{208103} (\bibinfo{year}{2007}).

\bibitem[{\citenamefont{Conti and
  {MacKintosh}}(2009)}]{conti_cross-linked_2009}
\bibinfo{author}{\bibfnamefont{E.}~\bibnamefont{Conti}} \bibnamefont{and}
  \bibinfo{author}{\bibfnamefont{F.~C.} \bibnamefont{{MacKintosh}}},
  \bibinfo{journal}{Physical Review Letters} \textbf{\bibinfo{volume}{102}},
  \bibinfo{pages}{088102} (\bibinfo{year}{2009}).

\bibitem[{\citenamefont{{MacKintosh} et~al.}(1995)\citenamefont{{MacKintosh},
  Kas, and Janmey}}]{mackintosh_elasticity_1995}
\bibinfo{author}{\bibfnamefont{F.~C.} \bibnamefont{{MacKintosh}}},
  \bibinfo{author}{\bibfnamefont{J.}~\bibnamefont{Kas}}, \bibnamefont{and}
  \bibinfo{author}{\bibfnamefont{P.~A.} \bibnamefont{Janmey}},
  \bibinfo{journal}{Physical Review Letters} \textbf{\bibinfo{volume}{75}},
  \bibinfo{pages}{4425} (\bibinfo{year}{1995}).

\bibitem[{\citenamefont{Vetterling and
  Flannery}(2002)}]{vetterling_numerical_2002}
\bibinfo{author}{\bibfnamefont{W.~T.} \bibnamefont{Vetterling}}
  \bibnamefont{and} \bibinfo{author}{\bibfnamefont{B.~P.}
  \bibnamefont{Flannery}}, \emph{\bibinfo{title}{Numerical Recipes in C++: The
  Art of Scientific Computing}} (\bibinfo{publisher}{Cambridge University
  Press}, \bibinfo{year}{2002}), \bibinfo{edition}{2nd} ed., ISBN
  \bibinfo{isbn}{0521750334}.

\bibitem[{\citenamefont{Broedersz et~al.}(2009)\citenamefont{Broedersz, Storm,
  and {MacKintosh}}}]{broedersz_effective-medium_2009}
\bibinfo{author}{\bibfnamefont{C.~P.} \bibnamefont{Broedersz}},
  \bibinfo{author}{\bibfnamefont{C.}~\bibnamefont{Storm}}, \bibnamefont{and}
  \bibinfo{author}{\bibfnamefont{F.~C.} \bibnamefont{{MacKintosh}}},
  \bibinfo{journal}{Physical Review E} \textbf{\bibinfo{volume}{79}},
  \bibinfo{pages}{061914} (\bibinfo{year}{2009}).

\bibitem[{\citenamefont{Broedersz and
  {MacKintosh}}(2010)}]{broedersz_future_2010}
\bibinfo{author}{\bibfnamefont{C.~P.} \bibnamefont{Broedersz}}
  \bibnamefont{and} \bibinfo{author}{\bibfnamefont{F.~C.}
  \bibnamefont{{MacKintosh}}}, \bibinfo{journal}{(to appear)}
  (\bibinfo{year}{2010}).

\end{thebibliography}

\end{document}